\documentclass[preprint,12pt]{elsarticle}

\usepackage{graphics}

\usepackage{amssymb}

\journal{Nuclear Instruments and Methods A}

\begin{document}

\begin{frontmatter}



\title{Rayleigh scattering and depolarization ratio in linear alkylbenzene}


\author[ucas]{Qian Liu}
\author[whu]{Xiang Zhou}
\author[ucas,whu]{Wenqian Huang}
\author[ucas]{Yuning Zhang}
\author[whu]{Wenjie Wu}
\author[ucas,whu]{Wentai Luo}
\author[whu]{Miao Yu}
\author[ucas]{Yangheng Zheng}
\author[ihep]{Li Zhou}
\author[ihep]{Jun Cao}
\author[ihep]{Yifang Wang}

\address[ucas]{University of Chinese Academy of Sciences,100049, Beijing, China}
\address[whu]{Hubei Nuclear Solid Physics Key Laboratory, Key Laboratory of Artificial Micro- and Nano-structures of Ministry of Education, and School of Physics and Technology, Wuhan University, Wuhan 430072, China}
\address[ihep]{Institute of High Energy Physics, Chinese Academy of Science, 100049, Beijing, China}

\begin{abstract}

Linear alkylbenzene (LAB) is adopted to be the organic solvent for the Jiangmen Underground Neutrino Observatory (JUNO) liquid scintillator detectors due to the ultra-transparency. However the current Rayleigh scattering length calculation disagrees with the measurement. The present paper for the first time reports the Rayleigh scattering of LAB being anisotropic and the depolarization ratio being $0.31\pm0.01(\mbox{stat.})\pm0.01(\mbox{sys.})$. We proposed an indirectly method for Rayleigh scattering measurement with  Einstein-Smoluchowski-Cabannes formula, and the Rayleigh scattering length of LAB is determined to be $28.2\pm1.0$\,m at $430$\,nm. 

\end{abstract}


\end{frontmatter}


\section{Introduction}

The Jiangmen Underground Neutrino Observatory (JUNO)  is a multipurpose neutrino experiment designed to determine neutrino mass hierarchy and precisely measure oscillation parameters by the medium baseline vacuum oscillations of the reactor antineutrinos \cite{juno1,juno2}. It's located in the $700$\,m deep underground laboratory at Jiangmen, China.

The design of JUNO is based on the principle with highly purified liquid scintillator (LS) as the central detector surrounded by ten-thousands of photomultiplier tubes (PMTs) and tons of ultra-pure water outside as an external shield. The central detector is a  20,000 tons of linear alkylbenzene (LAB) based LS in a spherical vessel with a diameter of 34.5 m, and the energy resolution is designed to be $3\%/\sqrt{E(\mbox{MeV})}$ corresponding to $1,200$ photoelectron (p.e.) collected by PMTs per MeV. The antineutrino is detected via the inverse beta decay reaction $\bar\nu_e+p\rightarrow e^+ + n$. LAB scintillation light generated from positron annihilation and neutron capture is red-shifted by the primary and secondary wavelength shifter \cite{shifter}, 2,5-diphenyloxazole (PPO) and pbis[2-methylstyryl]benzene (bis-MSB), and has to transverse through the whole LS vessel before arriving at the photomultipliers (PMTs). As a consequence, the transparency of LAB to the shifted scintillation light ($430$\,nm for instance) is essential \cite{energytotrans}. This has brought a great challenge to both of technical LAB purification and optical parameter measurement. 

The event location and energy reconstruction are determined by the optical modeling which includes Rayleigh scattering length, absorption length and attention length \cite{bybrec}. However, the absorption length which describes the energy of photon being absorbed to heat is difficult to measure directly. An indirectly method calculating from the attenuation length and scattering length is proposed\cite{Lequation}. Currently the attenuation length of LAB at $430$\,nm has been measured\cite{att1, att2, att3}, while the scattering length has a difference from $40$\,m \cite{wurm} to calculated $30$\,m \cite{zhouxandwurm}. These motivate a precise measurement of LAB Rayleigh scattering length at the scintillation wavelength of LS.

\section{Rayleigh scattering} \label{sec:theory}

Rayleigh scattering, developed by Rayleigh in 1899 \cite{rayleigh}, describes the light elastically scattering  off the medium molecules. For gas state, this theory was successfully applied to independently isotropic molecule, and modified by Cabannes with introducing a depolarization ratio to describe the anisotropy of molecules. For liquid state, due to the strong interaction effects between molecules, Einstein and Smoluchowski proposed scattering to be caused by the random motion of molecules which leads the fluctuations of density and the dielectric constant. The Rayleigh length of liquids can be described by Einstein-Smoluchowski-Cabannes (ESC) formula \cite{zhouxandwurm}:
\begin{equation}
l_{Ray} = \left\{\frac{8\pi^3}{3\lambda^4}\left[\frac{(n^2-1)(2n^2+0.8n)}{n^2+0.8n+1}\right]^2kT\beta_T\frac{6+3\delta}{6-7\delta}\right\}^{-1}.
\end{equation}
Here $\lambda$ is the wavelength of scattered light, $n$ is the refractive index, $k$ is the Boltzmann constant, $T$ the absolute temperature, $\beta_T$ the isothermal compressibility and $\delta$ is the depolarization ratio. For JUNO experiment, the temperature of the LS detector will be controlled at $T=20\pm1^{\circ}$C and the quantum efficiency of PMTs are optimized at $\lambda=430$\,nm. Recently, the $\beta_T$ of LAB at three temperatures over $4$ to $23^{\circ}$C has been measured by the vibrating tube method \cite{betaT}.  The refractive index $n$ of LAB from the same batch in the range between $400$\,nm and $630$\,nm has been reported \cite{zhouxandwurm}.

The depolarization ratio $\delta$ can be measured at $\theta=90^{\circ}$ scattering angle with a vertically polarized incident beam according to \cite{zhouxandwurm, morel}:
\begin{equation}
\delta_{90^\circ} = \frac{2I_h}{I_h+I_v}.
\end{equation}
The subscripts designate the components analyzed in the scattered beam. $v$ and $h$ are vertically and horizontally polarized portion, respectively. If we define depolarization ratio fraction $f=I_h/I_v$, we can get $\delta=2f/(f+1)$. 

\section{Experimental setup}

\begin{figure}[Experiment-Sch]
\centering
\includegraphics[width=.8\textwidth]{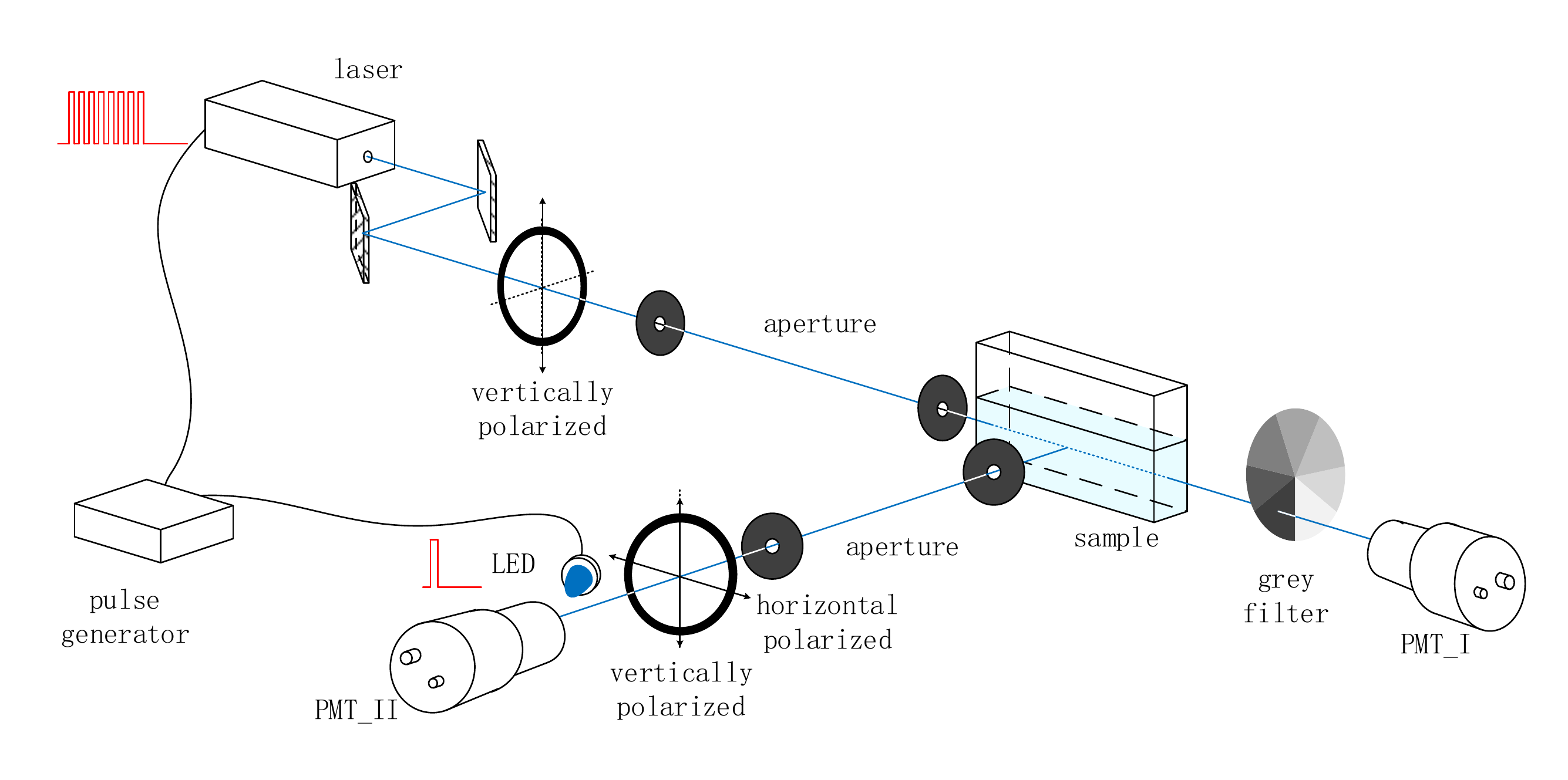}
\caption{Sketch of the experimental set-up}
\label{fig:Experiment-Sch}
\end{figure}

This experiment is designed to measure the depolarization ratio $\delta_{90^\circ}$ . The schematic setup is shown in Fig. \ref{fig:Experiment-Sch}. The light source is a Pico-Quant LDH pulsed laser diode with adjustable output up to $10$\,mW at $40$\,MHz repetition rate. Its wavelength is $405$\,nm instead of $430$\,nm because it's the only available pulsed laser we have which is the closest one to $430$\,nm. The beam from the laser has a divergence of $0.32$\,mRad, producing a beam diameter at the sample cell less than $1$\,mm. The incident beam is collimated with two variable apertures, as well as vertically polarized (polarization$<1\%$) with a Glan-Thompson polarizer. Fluctuations in the intensity of the light source are monitored by use of a reference PMT-I (Hamamatsu R2083) with a grey filter in front to block a portion of the incident light.  The samples are held in a quartz cuvette with 5-cm path lengths. The scattered light from the sample is then polarized by a Glan-Laser calcite polarizer with $100,000:1$ extinction ratio which is installed on a motorized rotation mount with $1''$ precision. For this measurement the selected polarization component is either vertically or horizontally to the plate defined by incident beam and scattered light. A Hamamatsu R1828 PMT-II with single photon response capability is installed after to count the number of scattered photons.  The performance of PMT-II is monitored with a blue LED. Both laser diode and LED are trigged by a pulse generator. This experiment is held in a dark room with room temperature controlled at 23\,$^\circ$C.

The signals from both PMTs are recorded by a CAEN DT$5720$ waveform digitizer with $4$\,nm time resolution ($250$\,MS/s) and a total range of $2$\,V at a $12$\,bit resolution. The trigger of data acquisition (DAQ) system is shared from the same pulse generator, and it's set to be $1$\,kHz with a $4\,\mu s$ time window. The trigger for laser diode is set to a burst mode, $60$ pulses with $5$\,ns time width are sent in this time window. Both of the signal from PMTs (usually PMT signals are less than $20$\,ns) and the thermal noise or radioactive background are recorded by DAQ. 

\section{Photoelectron counting} \label{sec:pc}

The intensity of scattered light is estimated by counting the number of p. e., then the depolarization ratio fraction $f$ can be written as: 
\begin{equation}
f=\frac{N_h}{N_v}=\frac{\sum\limits_{i=0}^{\infty} i \cdot N^i_h}{\sum\limits_{i=0}^{\infty}i \cdot N^i_v}.
\end{equation}
However, the horizontal and vertical portion of scattered light  are measured separately, which requires a careful consideration for normalizing the experimental conditions such as the stability of laser diode, DAQ data taking time period, etc. 

We propose a new method for estimating the fraction $f$. Assuming a Poisson distribution for the photon scattered and the p.e. number leaving the photocathode, and taking the number of incident photons to be $N$,  we can rewrite this to:
\begin{equation}
f=\frac{\sum\limits_{i=0}^{\infty} i \cdot \frac{N^i_h}{N}}{\sum\limits_{i=0}^{\infty}i \cdot \frac{N^i_v}{N}}=\frac{\mu_h}{\mu_v}.
\end{equation}
Here $\mu$ is the expected value for Poisson distribution.

Due to the long Rayleigh scattering length, the expected value $\mu$ would be rather small. The output of the PMT could be altered by the dark noise which causes a number of random coincidences are detected. Therefore, in order to have the random coincidences contribution at the level of $1\%$ it's necessary to have $\mu\geq f_{dark} \cdot \tau_{gate}/0.01$. Here $f_{dark}$ is the frequency of dark noise which is 4K for PMT-II, and $\tau_{gate}$ is the ADC gate which is $40$\,ns. It gives $\mu\geq 0.016$ to have a negligible contribution of the dark noise spectrum. 

For a smaller $\mu<0.016$, the observed Poisson distribution includes two parts. First one is the scattered photon response contribution, and the second one is the dark noise of PMT coming from the thermionic emission from the photocathode or radioactive background. The latter can be assumed to be a Poisson distribution as well. These two Poisson distributions are not correlated. Thus the fraction $f$ is:
\begin{equation} \label{eq:fraction}
f=\frac{\mu^{obs}_h-\mu^{dk}_h}{\mu^{obs}_v-\mu^{dk}_v}.
\end{equation}
The superscripts mean the observed and the dark noise background Poisson distribution. Since the PMT dark noise spectrum is not related to the polarization angle, the $\mu^{bg}_h$ and $\mu^{bg}_v$ should be at the same level.

To estimate the expected value $\mu$, the PMT charge spectrum is fitted with the following function: 
\begin{equation}
f(x)=N_0 \cdot \left[ P(0;\mu) \cdot f_{noise}(x) + P(1;\mu) \cdot f_{spe}(x)  + \sum_{n=2}^{N_{max}} P(n;\mu) \cdot f^n_{mpe}(x)\right].
\end{equation}
Here $N_0$ is a normalization factor, $P(n;\mu)$ is the probability of Poisson distribution with mean value $\mu$ for different contribution of $n$ photons. The description of the electronics noise function $f_{noise}(x)$, single photon response function $f_{sep}(x)$ and multiple photon response function $f_{me}^n(x)$ can be found somewhere at \cite{photocount}.

\begin{figure}[fig:SER]
\centering
\includegraphics[width=.45\textwidth]{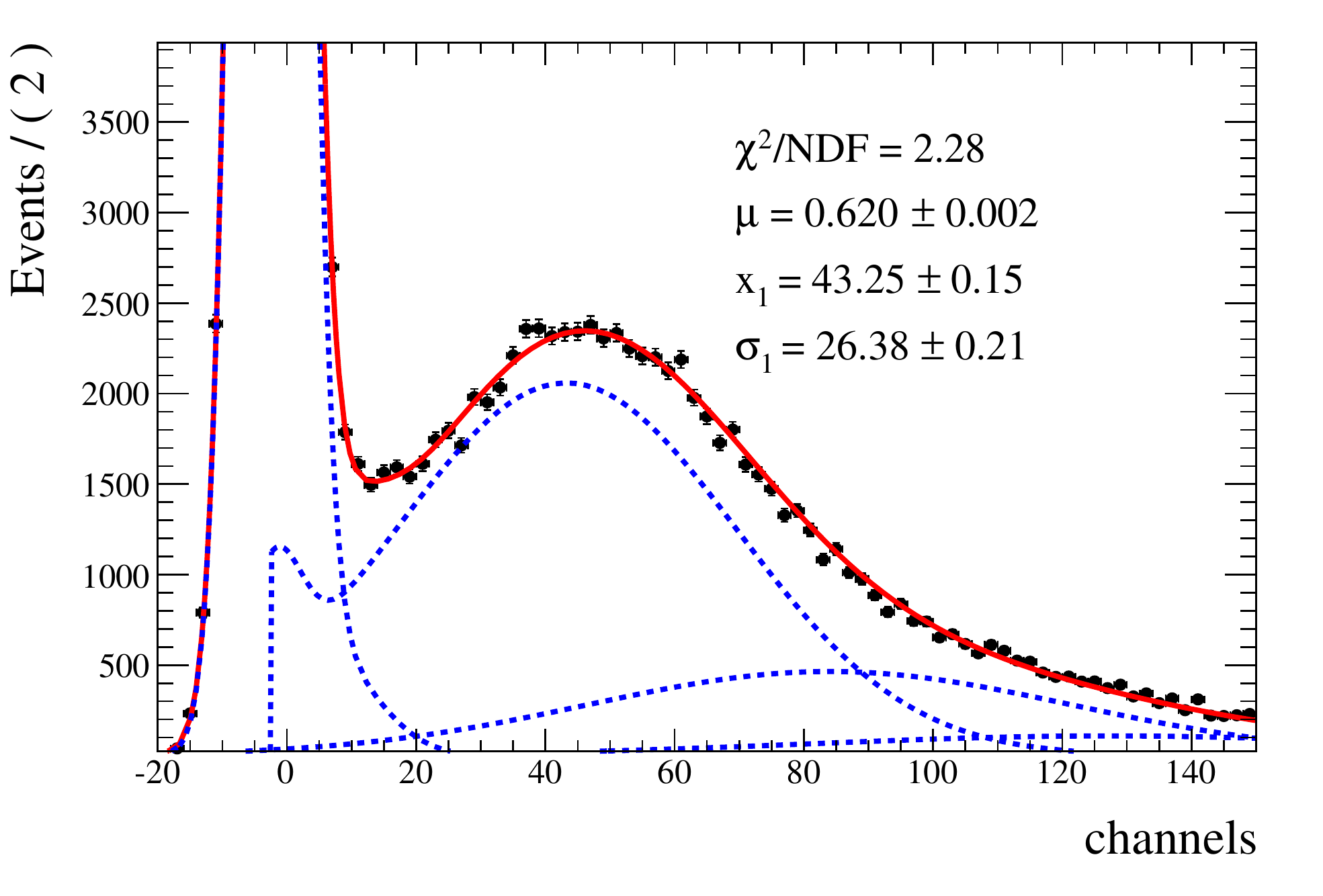}
\includegraphics[width=.45\textwidth]{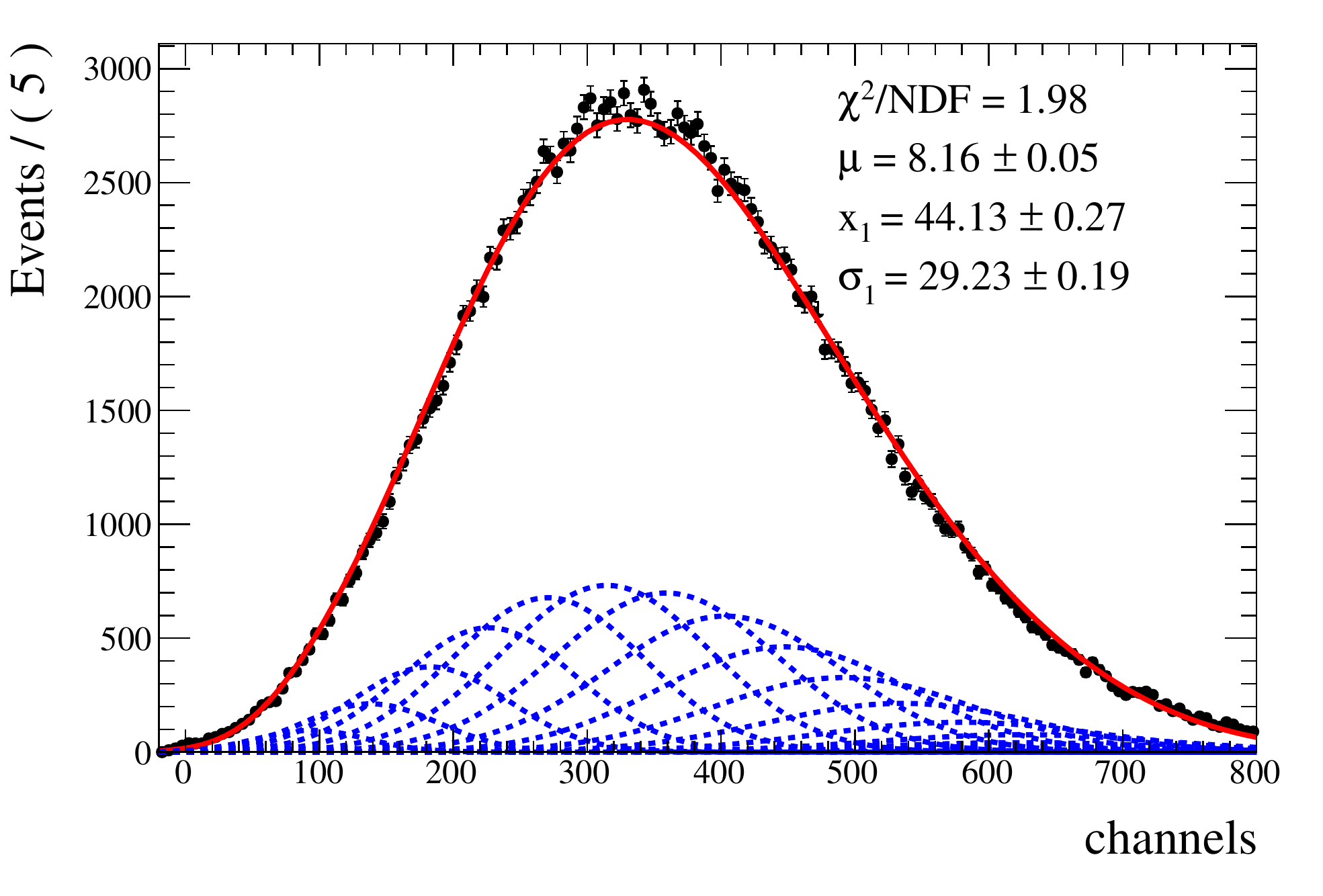}
\caption{The charge spectrum taken with two different mean photoelectron number. The dots are data, the red line is the fit result, and dashed lines are the contributions of each component.}
\label{fig:SER} 
\end{figure}

The charge response of the PMT-II for a low intensity light has been studied. The different $\mu$ of p.e. is achieved by tuning the applied voltage on the LED shown in Fig.\ref{fig:Experiment-Sch}. The charge spectrum taken with two different mean p.e. number are given in Fig.\ref{fig:SER}. On the left plot the single photon response including an exponential part and a Gaussian one is shown. Here the exponential part is to describe the thermionic emission contribution with a smaller energy or under unfavorable angles of incidence. The Gaussian part is the single p.e. response with the mean value $x_1$ and the standard deviation $\sigma_1$. The contribution of 2 and 3 p.e. can also be seen. On the right plot the multiple-photon response is given. 

\section{Result}

\begin{figure}[Experiment-Res]
\centering
\includegraphics[width=.45\textwidth]{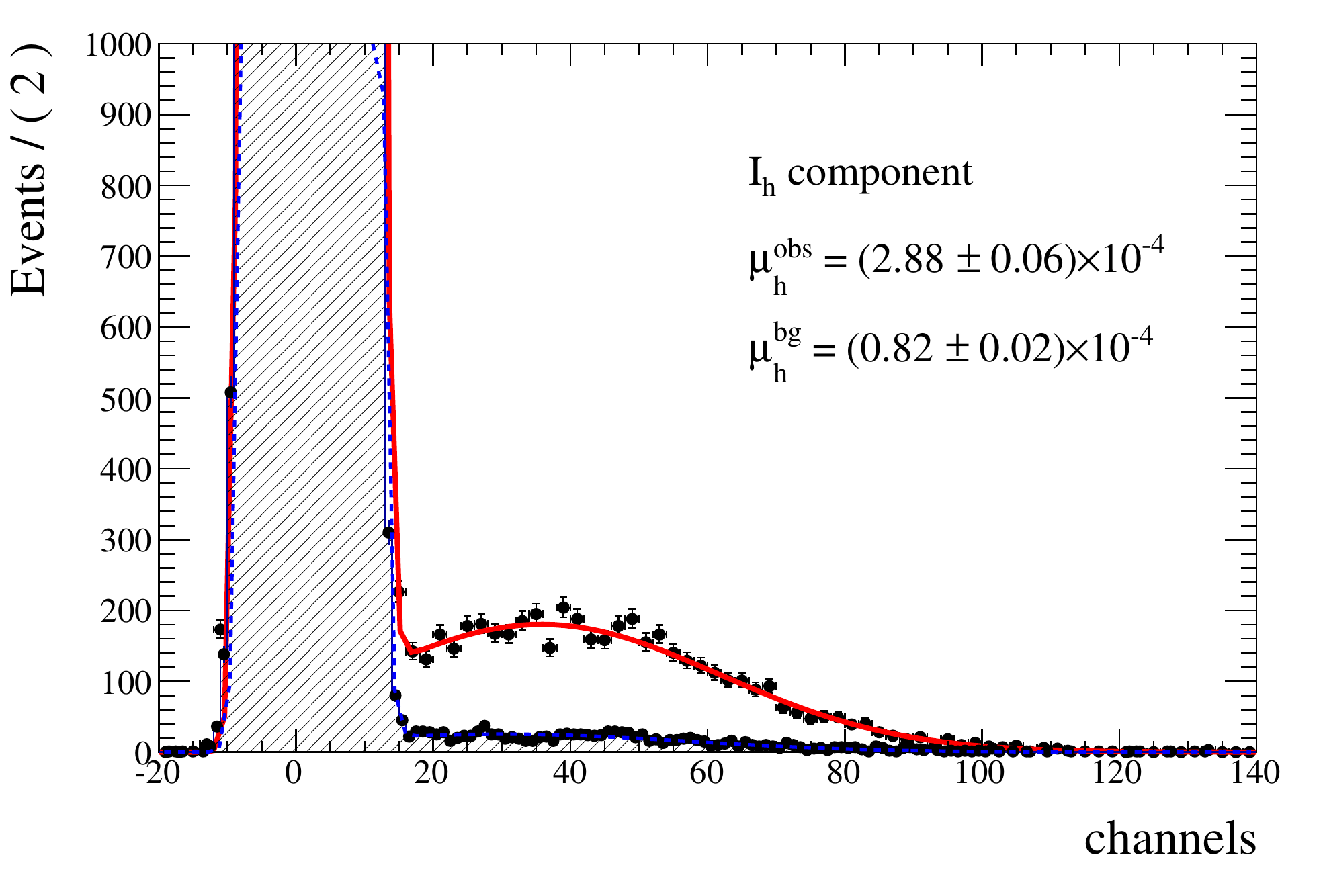}
\includegraphics[width=.45\textwidth]{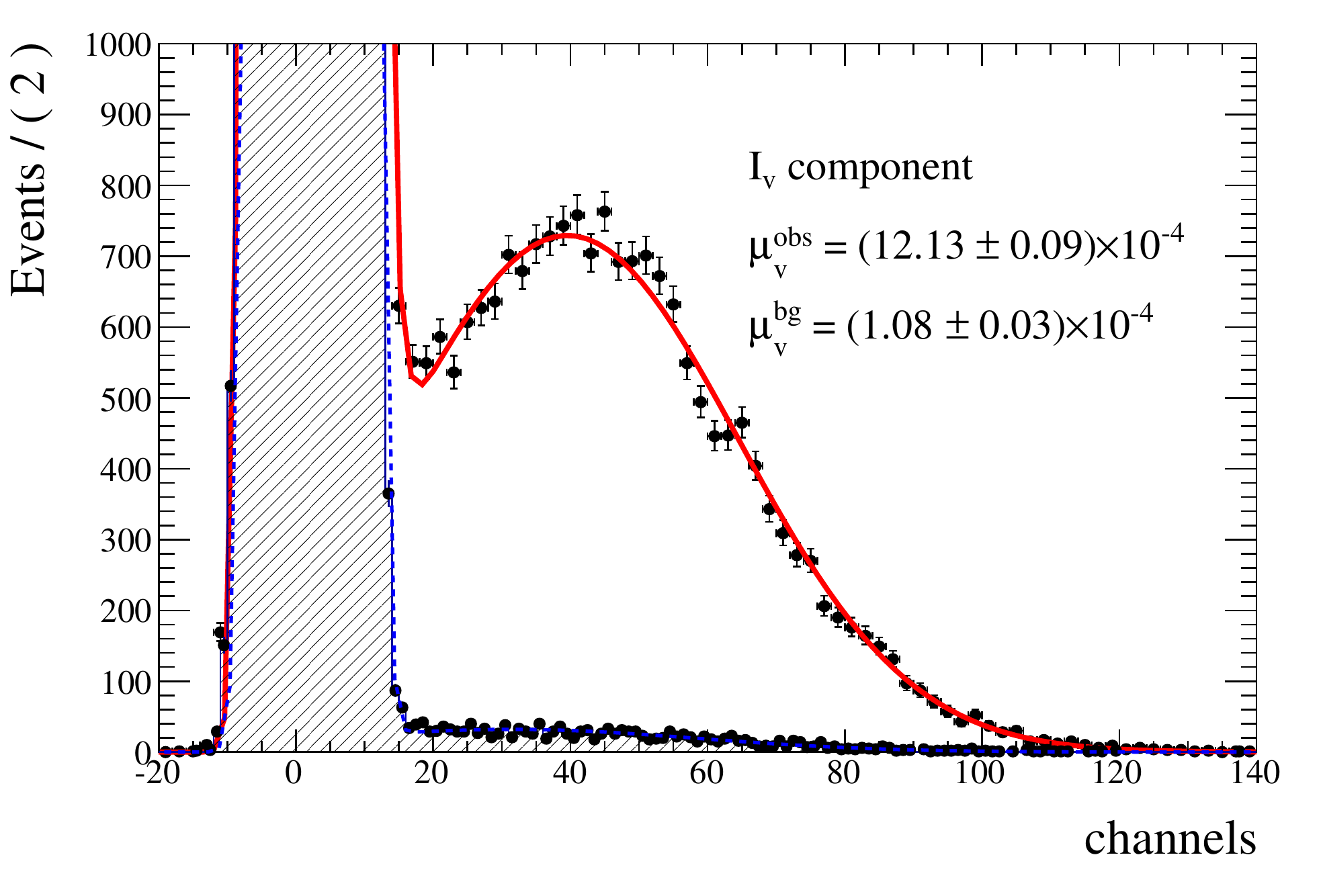}
\caption{The charge spectrum for vertical (left) and horizontal (right) portion of scattered light. The dots are data and red line is fitted spectrum. The shaded histogram is the dark noise spectrum and the dashed line is the fit result. }
\label{fig:Experiment-Res} 
\end{figure}

The results reported here include two samples. The first one is LAB, and the PMT spectrum is shown in Fig.\ref{fig:Experiment-Res}. The dots are observed data spectrum and the red line is fitted spectrum. The shaded histogram is the dark noise spectrum and the dashed line is the fit result. The measured depolarization ratio fraction is $f_{{LAB}}=0.186\pm0.006(\mbox{stat.})$, corresponding to $\delta_{{LAB}}=0.314\pm0.014(\mbox{stat.})$.

The second one is benzene with $99.5\%$ purity. As one of the most common organic solvents, benzene is deleterious and highly volatile. Careful consideration has been done during this measurement. The depolarization ratio fraction is measured to be $f_{Ben}=0.287\pm0.004(\mbox{stat.})$, which gives $\delta_{Ben}=0.44\pm0.01(\mbox{stat.})$. This is agreed within 3 $\sigma$ with the measurement made by Mossoulier which is about $0.42$ \cite{benzene}.

\section{Systematic error}

The systematic uncertainty of this measurement has been checked from the following  respects:

The wavelength of incident laser beam has been measured with Ocean Optics $4400$ spectrometer, and Gaussian fitted results show the wavelength of laser is $405\pm0.66$\,nm. The systematic error from this contribution is negligible.

The stability of laser is monitored with PMT-I. The intensity fluctuation of the incident laser beam is about $0.1\%$. As discussed in section \ref{sec:pc}, the expected value $\mu$ is not related to the intensity of the incident laser, this fluctuation will not be taken into account for the systematic error estimation.

During the measurement procedure, the cuvette will be washed and air dried for couple times. Even though the light path will be adjusted again after cleansing,  the position of the cuvette will still be slightly different. The systematic from this difference is considered by measuring the absorption spectrum of cuvette with a ultra-violet spectrophotometer (UV-Vis). By changing the position of cuvette in the UV-Vis, the difference of the absorption value at $405$\,nm is about $1.7\%$ and this is taken as the systematic error.

The acceptance correction for PMT-II is estimated by introducing a slit function. This slit function $S(\theta)$ can be approximately written as a function of the inner open-angle $\alpha$ and the outer open-angle $\beta$ \cite{slitfunction}.  In our measurement $\alpha=0.0598^\circ$, and $\beta=0.145^\circ$, this correction is estimated by calculating the integral $\int_{90-\beta/2}^{90+\beta/2}S(\theta)\cdot[{\delta-\bar\delta_{90}}]/{\bar\delta_{90}} \cdot d(\theta)$, and it shows a negligible contribution from this correction.

\begin{equation}
S(\theta)=
\left\{
\begin{array}{rcl}
1+\frac{\theta-(90-\alpha/2)}{(\beta-\alpha)/2} &  {[90-{\beta}/{2}, 90-{\alpha}/{2}]} \\
1 &   {[90-{\alpha}/{2}, 90+{\alpha}/{2}]} \\
1-\frac{\theta-(90+\alpha/2)}{(\beta-\alpha)/2} &  {[90+{\alpha}/{2}, 90+{\beta}/{2}]} \\
0 &  {\mbox{elsewhere}} \\
\end{array}.
\right.
\end{equation}

The precision of polarization angle is guaranteed by a high precision rotation mount with $\pm0.1^\circ$ resolution. For the polarization of the incident beam, this systematic error is estimated by rotating the polarizer from $-1^\circ$ to $1^\circ$, and the difference is about $2.2\%$. For the polarization of the scattered light, the systematic error is estimated by taking this uncertainty to the depolarization ratio calculation, and the largest offset is about $1.0\%$ which is taken as the systematic error from this contribution. 

The total systematic error is about $2.8\%$, and the details are given in Table.\ref{table:system}.

\begin{table}
\caption{Systematic error}
\centering
\begin{tabular}{lll}
\hline
\hline
Laser wavelength & negligible  \\
Polarization of incident beam & $2.2\%$ \\
Uniformity of cuvette & $1.7\%$ \\
Acceptance & negligible \\
Polarization of scattered light & $1.0\%$ \\
\hline
Total & $3.0\%$  \\
\hline
\end{tabular}
\label{table:system}
\end{table}

\section{Discussion and conclusion}

The depolarization ratio $\delta_{LAB}$ at $405$\,nm is measured to be $\delta_{LAB}=0.31\pm0.01(\mbox{stat.})\pm0.01(\mbox{sys.})$, here the first one is statistical error and the second one is systematic error. According to the benzene measurement\cite{benzene}, $\delta_{Ben}$ has a negligible difference between $405$\,nm and $430$\,nm. Assuming $\delta_{LAB}$ has the same value at $430$\,nm and taking the refractivity\cite{zhouxandwurm} and isothermal compressibility\cite{betaT}, the Rayleigh scattering length of LAB is about $28.2\pm1.0$\,m at $430$\,nm at room temperature. Here the LAB sample used in these measurements are purified from a set of steps including distillation, Al$_2$O$_3$ sorption, etc. The same purification technique will be used for the JUNO LAB mass production. 

The attenuation length of the same LAB sample at $430$\,nm has been measured to be about $19.44\pm0.62$\,m\cite{att3}. Thus the absorption length is $62.5\pm7.9$\,m, here the error is calculated from error propagation. The simulation shows that the energy resolution can reach the design goal with the parameters given above\cite{simulation}. 

Rayleigh scattering length, as one of the important optical parameters, describes the probability of photons scattered after transmitting a distance in medium. The shorter Rayleigh scattering length is, the higher probability photon being scattered. This increases the complexity of event vertex and energy reconstruction. To simplify this problem, it's better to have a long Rayleigh scattering. Our measurement of Rayleigh scattering length is closer to the calculated value reported in \cite{zhouxandwurm} than the direct measurement in \cite{wurm}. 

\section*{ACKNOWLEDGEMENTS}

This work has been supported by the Strategic Priority Research Program of the Chinese Academy of Sciences (Grant No.XDA10010500) and the Major Program of the National Natural Science Foundation of China (Grant No. 11390381).








\end{document}